 \documentstyle[12pt]{article}  

\let\a=\alpha \let\b=\beta \let\g=\gamma  
    
    \let\p=\pi

\let\la=\label  
   
\def\nn{\nonumber} \def\bd{\begin{document}} \def\ed{\end{document}}
\def\ds{\documentstyle} \let\fr=\frac \let\bl=\bigl \let\br=\bigr
\let\Br=\Bigr \let\Bl=\Bigl 
\let\bm=\bibitem
\let\na=\nabla
\let\pa=\partial \let\ov=\overline 
\def\ba{\begin{array}}
\def\ea{\end{array}}
\def\nn{\nonumber}
\newcommand{\ho}[1]{$\, ^{#1}$}
\newcommand{\hoch}[1]{$\, ^{#1}$}
\newcommand{\bea}{\begin{eqnarray}} 
\newcommand{\eea}{\end{eqnarray}} 
\newcommand{\ra}{\rightarrow}
\newcommand{\lra}{\longrightarrow}
\newcommand{\Lra}{\Leftrightarrow}
\newcommand{\ap}{\alpha^\prime}
\newcommand{\bp}{\beta^\prime}
\newcommand{\tr}{{\rm tr} }
\newcommand{\Tr}{{\rm Tr} } 
\newcommand{\NP}{Nucl. Phys. }
\newcommand{\tamphys}{\it\hoch{a} Center for Theoretical Physics,
Physics Department, Texas A \& M University, College Station, Texas
77843} 
\newcommand{\cern}{\it\hoch{d} Theory Division, CERN, CH--1211,Geneva
23}  
\newcommand{\mcgill}{\it\hoch{c} Physics Department, McGill
University, Montreal, PQ, H3A 2T8 Canada}
\newcommand{\damtp}{\it\hoch{b} DAMTP, University of Cambridge,
Silver St., Cambridge CB3 9EW, UK}
\newcommand{\auth}{M.~J.~Duff$^a$
\footnote{Research supported in part by NSF Grant PHY-99411543.},
Jonathan M.~Evans$^b$, Ramzi R.~Khuri$^c$
\footnote{Research supported by NSERC of Canada and Fonds FCAR du Qu\'ebec.}, 
J. X. Lu$^a$ 
\newline
and Ruben Minasian$^d$} 
\def\ct#1{\cite{#1}}
\def\req#1{(\ref{#1})}
\def\be{\begin{equation}}
\def\ee{\end{equation}}
\def\tr{\hbox{\rm tr}\,}
\def\Tr{\hbox{\rm Tr}\,}
\def\diag{\hbox{\rm diag}\,}
\def\ap{\a^\prime}
\def\bp{\b^\prime}
\def\za#1{{[#1]}}
\def\eqname#1{\label{#1}}
\def\gapij{{1\over 4 \p\ap}\,\sqrt{-\g}\,\g^{ij}}
\renewcommand{\theequation}{\arabic{section}.\relax \arabic{equation}}
\begin{document}

\hfill{CTP-TAMU-25/96}

\hfill{CERN-TH/97-156}

\hfill{DAMTP/96-78}

\hfill{McGill/97-14}

\hfill{hep-th/9706124}

\vspace{20pt}

\begin{center}
{ \large {\bf  The Octonionic Membrane }}

\vspace{36pt}

\auth

\vspace{20pt}

{\tamphys}

{\damtp}

{\mcgill}

{\cern}
\vspace{44pt}

\makeatletter
\@addtoreset{equation}{section} 
\makeatother
\renewcommand{\theequation}{\thesection.\arabic{equation}}
\addtolength{\baselineskip}{0.3\baselineskip}

\underline{ABSTRACT}
\vspace{8pt}
\end{center}
We generalize the supermembrane solution of $D=11$ supergravity by 
permitting the $4$-form $G$ to be either self-dual or anti-self-dual in
the eight dimensions transverse to the membrane. 
After analyzing the supergravity field  
equations directly, and also discussing 
necessary conditions for unbroken supersymmetry, we focus on two 
specific, related solutions.
The self-dual
solution is not asymptotically flat. The anti-self-dual solution is
asymptotically flat, has finite mass per unit area and saturates the
same mass=charge Bogomolnyi bound as the usual supermembrane.
Nevertheless, neither solution preserves any supersymmetry. Both
solutions involve the octonionic structure constants but, perhaps
surprisingly, they are unrelated to the octonionic instanton $2$-form
$F$, for which $TrF \wedge  F$ is neither self-dual nor
anti-self-dual.     

\vfill \footnoterule
{\footnotesize \vskip -12pt }
\baselineskip=24pt

\pagebreak

\setcounter{page}{1}
\section{Introduction}
\la{intro}
The eleven-dimensional supermembrane \cite{Bergshoeffsezgintownsend} is
one of the cornerstones of $M$-theory \cite{Schwarzpower,DuffM,TownsendM}. The
usual supermembrane solution of $D=11$ supergravity \cite{Duffstelle} has
symmetry $P_3 \times SO(8)$ and preserves half of the spacetime supersymmetry.
The only non-vanishing components of the $4$-form $G_{MNPQ}$
($M,N, \ldots =0,1,\ldots,10$) are the 
$G_{012m}$ ($m,n, \ldots =3,4,\ldots,10$). In this paper we introduce
generalizations of this solution
which are obtained by permitting the $4$-form $G_{mnpq}$ to be
non-vanishing and either self-dual or anti-self-dual in the eight
dimensions transverse to the membrane.

With $G_{mnpq}$ non-zero, the $SO(8)$ transverse symmetry of the usual
supermembrane solution is necessarily broken to some subgroup
(since no $SO(8)$-invariant fourth-rank antisymmetric tensors exist).
There are many possible choices for $G_{mnpq}$, each with their own
symmetry properties. Here we investigate in detail 
two particularly natural
choices involving a certain constant $SO(7)$-invariant tensor 
which is related to the algebra of octonions.
As a result, both solutions have symmetry $P_3 \times SO(7)$.
The self-dual solution is not
asymptotically flat. The 
anti-self-dual solution is asymptotically flat,
has finite mass per unit area and 
saturates the same mass=charge Bogomolnyi bound
as the usual supermembrane. Nevertheless, neither solution preserves any
supersymmetry and neither is free of curvature singularities. 
Although both solutions implicitly involve the octonionic structure
constants,
perhaps surprisingly, neither is related to the
octonionic $SO(7)$ instanton
$2$-form $F$ 
\cite{Corrigan,Fairlie,Fubini}, for which $TrF \wedge  F$ is neither 
self-dual nor anti-self-dual.     

\section{The eleven-dimensional supermembrane}
\la{supermem}
The bosonic sector of eleven-dimensional supergravity is described by
the action
\be
I_{11}={1\over 2\kappa^2} \int d^{11}x\sqrt{-g}
\left( R -
{1\over 2\cdot 4!} G_{MNPQ}G^{MNPQ} \right) -
{1\over 12\kappa^2} \int C \wedge G \wedge G ,
\la{sugra}
\ee
where $g_{MN}$ ($M,N=0,1,\ldots,10$) is the metric,
$C_3$ is a three-form gauge field with four-form field-strength
$G_4=dC_3$, and $*$ denotes Hodge duality. The equations of motion are
\be
R_{MN}-{1\over 2}g_{MN} R={1\over 12}
\left(G_{MPQR}G_N{}^{PQR} -{1\over 8} g_{MN}
G_{PQRS}G^{PQRS}\right),
\la{einstein}
\ee
and
\be
d * G=-{1\over 2} G\wedge G.
\la{3form} 
\ee
It was shown in \cite{Duffstelle} that the supergravity action $I_{11}$
admits a fundamental membrane solution preserving half the spacetime
supersymmetries.  The solution for a single membrane invariant under
$P_3\times SO(8)$, where $P_3$ is the $d=3$ Poincar\'e group, is given by
\be
ds^2=e^{2C/3}\eta_{\mu\nu} dx^\mu dx^\nu + e^{-C/3} \delta_{mn} dy^m dy^n,
\la{metric}
\ee
and 
\be
C_{012}=\mp e^C,
\la{form}
\ee
with 
\be
e^{-C}=1+K/y^6,
\la{C}
\ee
where $\mu,\nu=0,1,2$ are indices in the ``membrane'' directions,
$m,n=1,2,\ldots,8$ are indices in the eight-dimensional space
transverse to the membrane, $y$ is the radial coordinate in
this transverse space, and $K$ is a constant.  

In order that the above membrane provide a solution at $y=0$, it was
necessary to introduce an explicit source term. The source is 
a supermembrane sigma-model action
\be
S_3=T_3\int d^3\xi \left(-{1\over 2} \sqrt{-\gamma}
\gamma^{ij} \partial_i X^M \partial_j X^N g_{MN} + {1\over 2}
\sqrt{-\gamma}
\mp {1\over 3!} \epsilon^{ijk} \partial_i X^M \partial_j X^N
\partial_k X^P
C_{MNP}\right)
\ee
where $T_3$ is the membrane tension, $\xi^i$  are the coordinates on the
membrane worldvolume and $\gamma_{ij}$ is the worldvolume metric.  The
net effect of the sigma-model source term is to add a delta-function to
the right-hand side of the pure supergravity equations of motion and for
the ansatz given above (and with
$X^{\mu}=\xi^{\mu},X^m=constant$) they then reduce to the single
equation 
\be
\delta^{mn} \partial_m \partial_n e^{-C} = -2\kappa^2T_3\delta^8 (y). 
\ee
This fixes the constant $K=\kappa^2 T_3/3\Omega_7$, where $\Omega_7$ is
the volume of the unit seven-sphere.  Although this takes care of the
delta function at $y=0$, the solution can be analytically continued to
the region $r=0$ where $r$ is a Schwarzschild-like coordinate defined
by $r^6=y^6+K$
\cite{Gibbonstownsend,Duffgibbonstownsend} and the solution still
exhibits a curvature singularity at $r=0$.  This singularity contrasts
with the fivebrane soliton solution \cite{Gueven} which is everywhere
nonsingular.  The mass per unit area of the membrane ${\cal M}_3$ is equal
to its tension:
\be
{\cal M}_3=T_3 .
\ee
This elementary solution of the supergravity equations coupled to a
supermembrane source carries a Noether ``electric'' charge 
\be
Q=\frac{1}{\sqrt{2}\kappa}\int_{S^7}(*G+{1\over 2}C\wedge G)=\sqrt{2}\kappa T_3.
\ee
Hence the solution saturates the Bogomol'nyi bound
\be
\sqrt{2}\kappa {\cal M}_3 \geq Q .
\la{bound}
\ee
This follows from the preservation of half the supersymmetries,
although the converse is not true \cite{Dufflupopesezgin} (as we shall
rediscover in section \ref{octonion}). It is also intimately linked
with the worldvolume kappa symmetry of the fundamental supermembrane
\cite{Bergshoeffsezgintownsend}.

\section{Membrane solutions with (anti-)self-dual 4-forms}

We now wish to generalize the membrane solution of section 2 by
allowing for a non-vanishing $G_{mnpq}$. We make the same
ansatz (\ref{metric}) for the metric and (\ref{form}) for the 
components $C_{012}$ 
but we allow the function $C$ and the transverse components 
$C_{mnp}$ to be as yet unspecified functions of
the transverse coordinates $y^m$. 
(The ``mixed'' components of $C_{MNP}$ are zero.)
The $012mnpqr$
components of (\ref{3form}) are then equivalent to  
\be \partial_m[e^C(G^{mnpq} \mp
\frac{1}{4!}\epsilon^{mnpqrstu}G_{rstu})] = 0, 
\la{formeq1}
\ee
while the remaining components give
\be
 \delta^{mn}\partial_m \partial_n e^{- C} = 
\mp \frac{1}{2.4!}G_{mnpq} *G_{mnpq}.
\la{formeq2}
\ee
Transverse indices are now understood to be raised and lowered with the flat 
metric $\delta_{mn}$ and $*$ denotes the Hodge dual in
eight-dimensional Euclidean space.
Turning to the Einstein equations (\ref{einstein}), 
we find that the $\mu \nu$
components are satisfied if 
\be
 \delta^{mn}\partial_m \partial_n e^{- C} = 
-\frac{1}{2.4!}G_{mnpq}G^{mnpq}
\la{single}
\ee
while the only additional content of the $mn$ components 
is the vanishing of the eight-dimensional stress tensor:
\be
G_{mpqr}G_n{}^{pqr}-\frac{1}{8}\delta_{mn}G_{pqrs}G^{pqrs} = 0.
\la{stress}
\ee

We now observe that equations (\ref{formeq1}) and (\ref{stress}) are satisfied 
automatically 
when the $4$-form $G$ is self-dual or anti-self-dual in the eight
dimensions transverse to the membrane.  
Furthermore, the remaining equations (\ref{formeq2}) 
and (\ref{single}) then coincide.
So we conclude that the complete set of supergravity field equations
is satisfied for any 3-form $C_{mnp}$ whose field strength obeys 
\be
G_{mnpq} = \pm {1 \over 4!} \epsilon_{mnpq rstu} G_{rstu} ,
\la{dual}
\ee
provided the function $C$, which determines the remaining fields
through (\ref{metric}) and (\ref{form}), is 
a solution of the single equation (\ref{single}).
Concrete examples of such solutions are given in section \ref{octonion}.

\section{Supersymmetry}
\la{susy}

The fact that a bosonic field configuration is annihilated by one or
more
supersymmetries is usually assumed to imply that it must also satisfy 
the equations of motion, 
although the converse is not true. 
Having discussed the structure of the equations of motion in
the last section, we will now
examine the complementary issue of supersymmetry.
Once again, we consider a general 
field configuration given
by (\ref{metric}) and (\ref{form}) 
with $C$ and $C_{mnp}$ functions only of the 
coordinates $y^m$.

The full $D=11$ supergravity theory involves the bosonic
action $I_{11}$ in (\ref{sugra}) coupled 
to a gravitino field $\psi_M$.   
Under a supersymmetry transformation with local parameter $\zeta$,
the variation of the gravitino is 
\be
\delta \psi_M = \nabla_M \zeta
- {1 \over 288} G_{PQRS}
(\Gamma_M{}^{PQRS} - 8 \delta^P_M \Gamma^{QRS} ) \zeta . 
\ee
In the standard fashion, it suffices to consider bosonic field
configurations which are supersymmetric
in the sense that the above variation vanishes for one or more choice
of $\zeta$.

Corresponding to the metric ansatz (\ref{metric}), 
we can decompose the supersymmetry
parameter in the form $\zeta = \epsilon \otimes \xi$ and the 
$D=11$ curved space gamma-matrices can be written 
\be
\Gamma_\mu = e^{C/3} (\gamma_\mu \otimes \gamma_9)
, \qquad
\Gamma_m = e^{-C/6} ( 1 \otimes \gamma_m ) , 
\ee
where $\gamma_\mu$ and $ \gamma_m$ are gamma-matrices for flat
$D=3$ Minkowski space and $D=8$ 
Euclidean space respectively, and $\gamma_9$ is the eight-dimensional
chirality operator.
Using this decomposition in conjunction with (\ref{metric}) and
(\ref{form}) we find
\bea
\delta \psi_\mu  = 
\partial_\mu \zeta - {1 \over 6} e^{C/2} \partial_m C 
(\gamma_\mu \otimes \gamma_m \gamma_9) \zeta \phantom{XXXXXXXXXXXXXXX}
\nonumber \\
\pm {1 \over 6} e^{C/2} \partial_m C (\gamma_\mu \otimes \gamma_m) \zeta
- {1 \over 288} 
e^{C} G_{mnpq} (\gamma_\mu \otimes \gamma_{mnpq} \gamma_9) \zeta 
\eea
and
\bea
\delta \psi_m =
\partial_m \zeta - {1 \over 12} \partial_n C (1 \otimes
\gamma_{mn} ) \zeta \phantom{XXXXXXXXXXXXXXXXXXX} \nonumber \\
\mp{1 \over 6} \partial_m C (1 \otimes \gamma_9) \zeta 
\pm {1 \over 12} \partial_n C 
(1 \otimes \gamma_{mn} \gamma_9) \zeta \phantom{XXXXXXXX} \nonumber \\
+{1 \over 24} e^{C/2} G_{mnpq} (1 \otimes \gamma_{npq}) \zeta
- {1 \over 288} 
e^{C/2} G_{pqrs} (1 \otimes \gamma_m \gamma_{pqrs} ) \zeta 
\eea
In each of these equations, the term in the first line on the
right-hand side comes from the
spin-connection, while the remaining terms come from the decomposition
of the antisymmetric tensor fields. 

In order to arrange that all components of the gravitino have zero
variation, we cancel terms with
the same gamma-matrix structure by setting
\be 
\zeta = e^{C/6} (\epsilon \otimes \xi^\pm) , \qquad \gamma_9 \xi^\pm = \pm
\xi^\pm,\la{chiral}
\ee
with $\epsilon$ and $\xi^\pm$ constant.
The condition for a supersymmetric field configuration 
then reduces to a single equation in the transverse space: 
\be
G_{mnpq} \gamma_{npq} \xi^\pm = 0.
\la{super}
\ee
(Related equations made their appearance in the physically 
different context of finding Calabi-Yau fourfold compactifications of 
$D=11$ supergravity down to $D=3$ \cite{Becker}.)

We emphasize that we have assumed nothing so far about 
$C_{mnp}$ or its field strength $G_{mnpq}$.
Note that if $G_{mnpq} = 0$ we recover the fundamental membrane solution of 
section 2, and (\ref{super}) then gives no additional restriction 
on $\xi^\pm$, implying that half of the spacetime
supersymmetries of $D=11$ supergravity are preserved.  
In the case with  $G_{mnpq}$ non-zero,
we would like to understand how the condition for
supersymmetry (assuming (\ref{chiral})) fits together with our
analysis of the equations of motion; or in other words,
whether (\ref{super}) can be satisfied 
when $G_{mnpq}$ is self-dual or anti-self-dual. 

It is natural to expect that there are no spinors $\xi^\pm$ which 
satisfy (\ref{super}) when 
$ *G = \mp G$. This is because the sign specifying the chirality of
the spinor in the condition for unbroken supersymmetry
is correlated with the sign appearing in the ansatz
(\ref{form}). But from section 3 we know 
that the sign in (\ref{form}) is in turn correlated with the sign in 
(\ref{dual})
if we are to have a solution of the field equations.
A spinor $\xi^\pm$ which satisfied (\ref{super}) with 
$ *G = \mp G$ would therefore imply the existence of a supersymmetric field
configuration which did not satisfy the equations of motion, which
runs counter to conventional wisdom. 
Indeed, it can be shown directly
from (\ref{super}) that no such configurations are possible.

To establish this, we first note that for any  
commuting spinors $\xi^\pm$ 
with $\gamma_9 \xi^\pm = \pm \xi^\pm$, the tensors 
\be 
X^\pm_{mnpq} = 
\xi^{\pm T} \gamma_{mnpq} \xi^\pm
\ee are self-dual and anti-self-dual:
\be X^\pm_{mnpq} = \pm 
{1 \over 4!} \epsilon^{mnpq rstu} X^{\pm}_{rstu} .
\ee
Now on squaring (\ref{super}) and performing some gamma-matrix algebra we find 
\be 
2 (\xi^\pm)^2 G_{mnpq} G_{mnpq} = 
3 G_{mn pq} G_{mn rs}  X^\pm_{pq rs } .
\ee
But it is also easy to show that $ *G = \mp G $ and $*X^\pm = \pm
X^\pm$ together imply that the expression on the right-hand side
must vanish, and hence that $G_{mnpq} = 0$.

In conclusion, we have shown that for a non-zero field-strength $G$ 
which is self-dual or anti-self-dual, the only 
unbroken supersymmetries allowed by
(\ref{super}) are given by spinors with positive or negative 
chirality respectively. 
Whether there actually are any such
unbroken supersymmetries of this type 
is then a question which depends on the detailed 
structure of $G$. 

\section{Octonions and related tensors}

We now turn from the general considerations of the previous sections
to discuss some specific solutions. 
Although no configuration with non-zero $G$ can be $SO(8)$-invariant,
we can find solutions with 
$SO(7)$ symmetry by making use of a certain
tensor $c_{mnpq}$ which is related to the algebra of octonions. 
We will begin by introducing this tensor from a slightly different
point of view, however.

Choosing any
commuting, positive-chirality spinor $\eta$ and normalizing it 
so that $\eta^T \eta = 1$, we define 
\be
c_{mnpq}=\eta^T\gamma_{mnpq} \eta.
\ee
As we mentioned in the last section,
this tensor is self-dual, obeying 
\be
c_{mnpq}=  \frac{1}{4!}\epsilon^{mnpqrstu}c_{rstu}.
\la{sconst}
\ee
Some other useful identities are
\be
c_{mnpr}c^{mnqs}=12\delta_{[pr]}{}^{[qs]}-4c_{pr}{}^{qs} , \quad
c_{mnpr}c^{mnps}=42\delta_r{}^{s} , \quad
c_{mnpq}c^{mnpq}=336 ,
\ee
which can also be deduced using standard properties of gamma-matrices.

It is immediate from its definition that the 
tensor $c_{mnpq}$ is invariant
under a maximal subgroup $SO(7) \subset SO(8)$
with respect to which the 
positive-chirality spinors decompose as 
$8 \to 7 \oplus 1$, with $\eta$ belonging to the singlet. 
The eight-dimensional vector and negative-chirality spinor 
representations remain irreducible under this subgroup.
Any two choices of the fixed spinor $\eta$ are equivalent, in that
they correspond to conjugate 
$SO(7)$ embeddings in $SO(8)$.  

To explain the relationship to octonions,
we introduce the totally antisymmetric 
octonionic structure constants $c_{abc}$ by means of the
multiplication 
rule 
\be
o_{a}o_{b}=-\delta_{ab} + c_{abc}o_{c},
\ee
where $o_{a}$ are unit imaginary octonions with 
$a,b,\ldots =1,2,\ldots ,7$.
With suitable choices of bases, the connection between $c_{mnpq}$
and $c_{abc}$ is simply
\be c_{abc8}=c_{abc},\qquad
c_{abcd}= {1\over 3!} \epsilon_{abcdefg} c_{efg} .
\la{const}
\ee
The description 
in terms of octonions is attractive from many points 
of view. It has one disadvantage, however, in that it makes manifest only a 
$G_2$ subgroup (the octonion automorphism group) of
the full $SO(7)$ symmetry of $c_{mnpq}$.

\section{An octonionic membrane}
\la{octonion}

Using the tensor $c_{mnpq}$ introduced in the last
section, we can construct both self-dual and anti-self-dual
solutions of the supergravity field equations which are 
invariant under $P_3 \times SO(7)$.

To find a self-dual $G$ we write
\be
C_{mnp}=\frac{1}{a}c_{mnpq}y^q,
\ee
where $a$ is a constant, and hence
\be
G_{mnpq}=\frac{1}{a}c_{mnpq}
\la{selfdual}
\ee
is manifestly self-dual. To find an anti-self-dual $G$ we write
\be
C_{mnp}=\frac{a^7}{y^8}c_{mnpq}y^q
\ee
and hence
\be
G_{mnpq}=-\frac{a^7}{y^{10}}(y^2c_{mnpq}+ 8 y_{[m}c_{npq]r}y^r),
\la{antiselfdual}
\ee
whose anti-self-duality follows from the self-duality of $c_{mnpq}$.
(In general, $c_{mnpq}$ self-dual and $h_{mn}$ symmetric and traceless
implies $c_{r[mnp}h_{q]r}$ is anti-self-dual.)

In the self-dual case, substituting into (\ref{single}) we find 
\be
\frac{1}{y^7}\frac{\partial}{\partial y}
            (y^7\frac{\partial}{\partial y}e^{-C})= - \frac{7}{a^2}
\ee
and hence
\be
e^{- C}=1+ \frac{K}{y^6} - \frac{7 y^2}{16 a^2},
\la{self} 
\ee
which yields a metric which is not asymptotically flat.  In the anti-self-dual 
case, we find
\be
\frac{1}{y^7}\frac{\partial}{\partial y}
            (y^7\frac{\partial}{\partial y}e^{-C})=-\frac{7a^{14}}{y^{16}}
\ee
and hence
\be
e^{- C}=1+ \frac{K}{y^6}-\frac{a^{14}}{16y^{14}},
\la{anti} 
\ee
which yields a metric which is asymptotically flat.
Note the minus signs in both (\ref{self}) and (\ref{anti}) which
imply that the metric (\ref{metric}) is singular at the zeros of
$e^{- C}$. In each case there is exactly one such zero at $y=y_0>0$.
These zeros actually give rise to curvature singularities.  

It is not difficult to read off the mass and charge of the new 
asymptotically flat, anti-self-dual
solution and we find the same results as for the usual supermembrane, in
spite of the non-vanishing of $G_{mnpq}$, essentially because the
coefficient of the $y^{-6}$ term in ($\ref{anti}$) remains unchanged.
The bound (\ref{bound}) is therefore still saturated by this new
solution.

From section \ref{susy}, we know that the only
possible unbroken supersymmetries for $*G = \pm G$ 
involve spinors $\xi^\pm$. 
For the specific $G_{mnpq}$ appearing in both (\ref{self}) and 
(\ref{anti}), we find that none of the supersymmetry
survives. This may seem surprising in view of the fact that the
anti-self-dual solution saturates the same Bogomolnyi bound  between
the mass and charge (\ref{bound}) as the usual supermembrane, however 
the relation between supersymmetry and the Bogomolnyi bound is a
subtle one 
\cite{Dufflupopesezgin} and can break down in the presence of singularities. 

By replacing 
$c_{mnpq}$ with other constant tensors it is clear that we can obtain
a variety of similar solutions with
invariance groups such as $SU(4) \times U(1)$ or $SO(4) \times SO(4)$ 
\cite{Corrigan}.
We do not expect any improvement in the singularity structure of
these solutions, however.
Furthermore, 
the $SO(7)$-invariant choice $c_{mnpq}$ is particularly
natural in conjunction with the isotropic metric ansatz (\ref{metric}), since
this is the unique maximal subgroup under which the 
vector representation of $SO(8)$ remains irreducible.
For any other subgroup, the condition that the metric
should depend only on the radial transverse coordinate would be much
less compelling.   
Allowing more general transverse behaviour also opens up 
many other possibilities 
of course, and we note in particular that
$G_{mnpq}$ is anti-self-dual whenever
\be
C_{mnp}=c_{mnpk} \partial_k f(y^i) , \qquad 
\delta^{mn} \partial_m \partial_n f=0 .
\la{antigen}
\ee
Our octonionic anti-self-dual solution utilizes 
the maximally symmetric harmonic function. Once again, 
we would not expect any less symmetric solution to have a more
desirable singularity structure. 

We should also mention that
having obtained a new membrane solution in $D=11$, it follows
by simultaneous dimensional reduction that we can obtain a new type
IIA string solution in $D=10$ \cite{Howe,Duffstelle}.

\section{Not the octonionic instanton}

The octonionic structure constants have appeared before in many 
different physical contexts 
\cite{Englert,Corrigan,DeWit,Fairlie,Fubini,Duffnilssonpope,Harveystrominger,Evans,%
Duffkhurilu,Ivanova,Gunaydin,Baulieu}. In particular, octonionic string
soliton 
\cite{Harveystrominger} and 
octonionic membrane soliton \cite{Ivanova,Gunaydin} solutions of the heterotic
 string have been found which make use of the $SO(7)$
octonionic instanton 
\cite{Corrigan,Fairlie,Fubini} in eight 
dimensions and the $G_{2}$ octonionic instanton \cite{Ivanova,Gunaydin} in seven
dimensions  respectively. 
In the Horava-Witten \cite{Horavawitten}
approach to deriving the $D=10$ $E_8 \times E_8$ heterotic string by compactifying
$M$-theory on $S^1/Z_2$, the equation 
\be 
G \sim Tr F^2 -\frac{1}{2}trR^2
\ee
appears on one of the boundaries, where $F$ is the Yang-Mills field strength of 
one of the $E_8$'s. One might therefore have expected that our solution
would be related to the $SO(7)$-invariant octonionic instanton where the tensor
$c_{mnpq}$ also makes its appearance in the equation
\be
F_{mn}=\frac{1}{2}c_{mnpq}F^{pq}.
\ee 
Indeed, for instanton size $\rho$ one finds \cite{Fairlie,Fubini}  
 \be
Tr F_{[mn}F_{pq]}= \frac{3{\rho}^2+y^2}
{({\rho}^2+y^2)^3}c_{mnpq} +
\frac{4(4{\rho}^2+y^2)}
{({\rho}^2+y^2)^4}y_{[m}c_{npq]r}y^r
\ee
which has a similar tensor structure as (\ref{antiselfdual}). However, it
is not difficult to verify that this expression is neither self-dual nor
anti-self-dual and does not provide a solution for $ G_{mnpq}$. In this
respect we differ from the authors of \cite{Baulieu} who claim that
$Tr F \wedge F$ is self-dual.

Another way in which our supermembrane solution differs from the
octonionic string and membrane solutions of
\cite{Harveystrominger,Ivanova,Gunaydin} is that the string has
infinite mass per unit length and the membrane has infinite mass per
unit area, whereas  our anti-self-dual solution has the same finite
mass per unit area as the usual supermembrane \cite{Duffstelle}. 

\section{Conclusions}

Although the octonionic instanton $4$-form $TrF^2$ is neither self-dual
nor anti-self-dual, we have not entirely given up on the possibility 
that Yang-Mills instantons may provide a solution to the $D=11$ supergravity 
equations. There are $M$-theoretic corrections to the $D=11$ supergravity
Lagrangian arising from a sigma-model Lorentz anomaly on the worldvolume of the
fivebrane \cite{Duffliuminasian}. The $3$-form field equation gets
modified to 
\be
d * G=-{1\over 2} G\wedge G +(2\pi)^{4}\beta X_{8},
\la{new3form} 
\ee
where $\beta$ is related to the fivebrane tension $T_{6}$ by 
$T_{6}=1/(2\pi)^{3}\beta$ and where
\be
X_{8}= \frac{1}{(2\pi)^{4}}[-\frac{1}{768}(tr R^{2})^{2} + 
\frac{1}{192} tr R^{4}].
\ee 
Indeed, our original motivation for generalizing
the supermembrane solution was to
search for solutions involving eight-dimensional Yang-Mills instantons
in the dimensions transverse to the membrane whose finite
instanton size would smear out the singularity at $r=0$ of the
usual supermembrane solution \cite{Duffstelle}. We also expected that
the $(R^2)^2$ and $R^4$ $M$-theoretic corrections (\ref{new3form}),
together with corresponding corrections to the Einstein equations
demanded by supersymmetry \cite{Greenvanhove}, would also play a
role.  One's first inclination might be to look for solutions of this kind
which preserve the $SO(8)$ symmetry of the usual supermembrane
solution. Indeed, $SO(8)$ Yang-Mills instantons do exist
\cite{Grossman} for which $F^2$ is in fact self-dual. However $TrF^2$
necessarily vanishes since there are no 
$SO(8)$-invariant antisymmetric tensors of
rank $4$. $TrF^4$ is non-zero, however, and indeed these instantons 
played a role in smoothing out \cite{Dufflu} the singularity of the $SO(32)$
heterotic string soliton solution \cite{Dabholkar} 
by incorporating the one-loop
$TrF^4$ corrections to the Lagrangian. 

We intend to return elsewhere to this original goal of finding
non-singular
supersymmetric solutions involving the $M$-theoretic corrections.  In this paper,
however, we were diverted into finding singular, non-supersymmetric solutions for
which the $X_8$ corrections vanish, but which may nevertheless prove to be of
interest in their own right. 

\section{Acknowledgements}

MJD is grateful for hospitality extended by the New High Energy Theory
Group, Rutgers University, and by the Isaac Newton Institute, Cambridge,
where part of this work was carried out. JME is supported by a PPARC
Advanced Fellowship. RM is supported by a World Laboratory Fellowship.

\end{document}